\documentclass[prl,twocolumn,showpacs,amsmath,amssymb,superscriptaddress]{revtex4-1}
\usepackage{dcolumn}
\usepackage{bm,graphicx}
\usepackage{etoolbox}
\usepackage{url}
\usepackage[colorlinks]{hyperref}
\usepackage{breakurl}
\usepackage{color}
\usepackage [latin1]{inputenc}

\begin{document}

\title{Magneto-optics of a Weyl semimetal beyond the conical \\ band approximation:  case study of TaP}

\author{S.~Polatkan}
    \affiliation{1.~Physikalisches Institut, Universit\"at Stuttgart, Pfaffenwaldring 57, 70569 Stuttgart, Germany}
\author{M.~O.~Goerbig}
    \affiliation{Laboratoire de Physique des Solides, Universit\'e Paris-Saclay, CNRS UMR 8502, 91405 Orsay Cedex, France}
\author{J.~Wyzula}
\affiliation{Laboratoire National des Champs Magn\'etiques Intenses, CNRS-UGA-UPS-INSA-EMFL, 25 rue des Martyrs, 38042 Grenoble, France}
\author{R.~Kemmler}
    \affiliation{1.~Physikalisches Institut, Universit\"at Stuttgart, Pfaffenwaldring 57, 70569 Stuttgart, Germany}
\author{L.~Z.~Maulana}
    \affiliation{1.~Physikalisches Institut, Universit\"at Stuttgart, Pfaffenwaldring 57, 70569 Stuttgart, Germany}
\author{B.~A.~Piot}
\affiliation{Laboratoire National des Champs Magn\'etiques Intenses, CNRS-UGA-UPS-INSA-EMFL, 25 rue des Martyrs, 38042 Grenoble, France}
\author{I.~Crassee}
\affiliation{Laboratoire National des Champs Magn\'etiques Intenses, CNRS-UGA-UPS-INSA-EMFL, 25 rue des Martyrs, 38042 Grenoble, France}
\author{A.~Akrap}
    \affiliation{Department of Physics, University of Fribourg, Chemin du Musée 3, CH-1700 Fribourg, Switzerland}
\author{C.~Shekhar}
\affiliation{Max Planck Institut f\"{u}r Chemische Physik fester Stoffe, 01187 Dresden, Germany}
\author{C.~Felser}
\affiliation{Max Planck Institut f\"{u}r Chemische Physik fester Stoffe, 01187 Dresden, Germany}
\author{M.~Dressel}
\affiliation{1.~Physikalisches Institut, Universit\"at Stuttgart, Pfaffenwaldring 57, 70569 Stuttgart, Germany}
\author{A.~V.~Pronin}
\affiliation{1.~Physikalisches Institut, Universit\"at Stuttgart, Pfaffenwaldring 57, 70569 Stuttgart, Germany}
\author{M.~Orlita}\email{milan.orlita@lncmi.cnrs.fr}
\affiliation{Laboratoire National des Champs Magn\'etiques Intenses, CNRS-UGA-UPS-INSA-EMFL, 25 rue des Martyrs, 38042 Grenoble, France}
\affiliation{Charles University, Faculty of Mathematics and Physics, Institute of Physics, Ke Karlovu 5, 121 16 Prague 2, Czech Republic}

\date{\today}

\begin{abstract}
Landau-level spectroscopy, the optical analysis of electrons in materials subject to a strong magnetic field, is a versatile probe of the electronic band structure and has
been successfully used in the identification of novel states of matter such as Dirac electrons, topological materials or Weyl semimetals. The latter arise from a
complex interplay between crystal symmetry, spin-orbit interaction and inverse ordering of electronic bands. Here, we report on unusual Landau-level transitions
in the monopnictide TaP that decrease in energy with increasing magnetic field. We show that these transitions arise naturally at intermediate energies in
time-reversal-invariant Weyl semimetals where the Weyl nodes are formed by a partially gapped nodal-loop in the band structure.
We propose a simple theoretical model for electronic bands in these Weyl materials that captures the collected magneto-optical data to great extent.
\end{abstract}

\maketitle

By now, the existence of Weyl semimetals is well-established, both by experimental facts~\cite{XuScience15,LvPRX15,LvNaturePhys15,YangNaturePhys15,XuNaturePhys15,HuangNatureComm15,LiuNM16} as well as by a solid theoretical basis~\cite{HerringPR37,BurkovPRL11,WanPRB11,WengPRX15,BansilRMP16,ChiuRMP16,ArmitageRMP18}. These are condensed-matter systems with electronic excitations described by the Weyl equation and lacking the degeneracy with respect to spin. Such conical bands may appear, due to accidental crossing of spin-split electronic bands, in any system with sufficiently strong spin-orbit coupling and with the lack of either space-inversion or time-reversal symmetry.

At present, the family of transition metal monopnictides (TaAs, TaP, NbAs and NbP) is likely the best-known and experimentally most explored
class of Weyl semimetals. These are non-magnetic type-I Weyl materials in which the non-degenerate conical bands appear due
to missing space inversion. Numerical simulations~\cite{HuangNC15,WengPRX15,LeePRB15,GrassanoJAP18,GrassanoSR18,GrassanoCM19} indicate that there are two kinds of Weyl nodes
nearby the Fermi energy in monopnictides, labeled usually as W$_1$ and W$_2$. The pairs of nodes with opposite chiralities (4 W$_1$ and 8 W$_2$ pairs in total) are
always oriented perpendicular to the tetragonal axis and distributed close to the boundaries of the first Brillouin zone.

In this work, we propose a theoretical model which allows us to describe the electrical and optical properties of time-reversal-invariant Weyl
semimetals. We test the model using a comparison with our magneto-optical experiments on TaP, a well-known Weyl semimetal from  the monopnictide family.
Our approach goes beyond the common approximation of a conical band~\cite{AshbyPRB13,TchoumakovPRL16,YuanNC18,JiangNL18,SinghPRB19}
and addresses the magneto-optical response due to specific features necessarily appearing in the band structure -- the band extrema due to the inversion and saddle
points of merging Weyl cones.

To describe massless excitations in a time-reversal-invariant Weyl semimetal, we start with a simple generic 2$\times$2 Hamiltonian often applied to systems
with inverted bands:
\begin{equation}
\hat{H} = \left( \begin{matrix}
\Delta - \frac{\hbar^2\mathbf{q}^2}{2M} & \gamma (\mathbf{q})\\
\gamma^* (\mathbf{q})& -\Delta + \frac{\hbar^2\mathbf{q}^2}{2M}
\end{matrix} \right)
\label{hamiltonian_Mark}
\end{equation}
that implies the electronic band structure, $E(\mathbf{q})=\pm \sqrt{[\Delta-\hbar^2\mathbf{q}^2/(2M)]^2+|\gamma(\mathbf{q})|^2}$, with
a full electron-hole symmetry.
The off-diagonal coupling term $\gamma(\mathbf{q})$ determines the nature of the described system, \emph{e.g.}, a gapless ($\gamma=0$) and a gapped ($\gamma = \mathrm{const.}$)
nodal-loop semimetal~\cite{PalPRB2016}, or a topological insulator, $\gamma(\mathbf{q})=\hbar v(q_x+iq_y)$~\cite{BernevigScience06,LiuPRB10}.

\begin{figure*}[t]
      \includegraphics[width=0.7\textwidth]{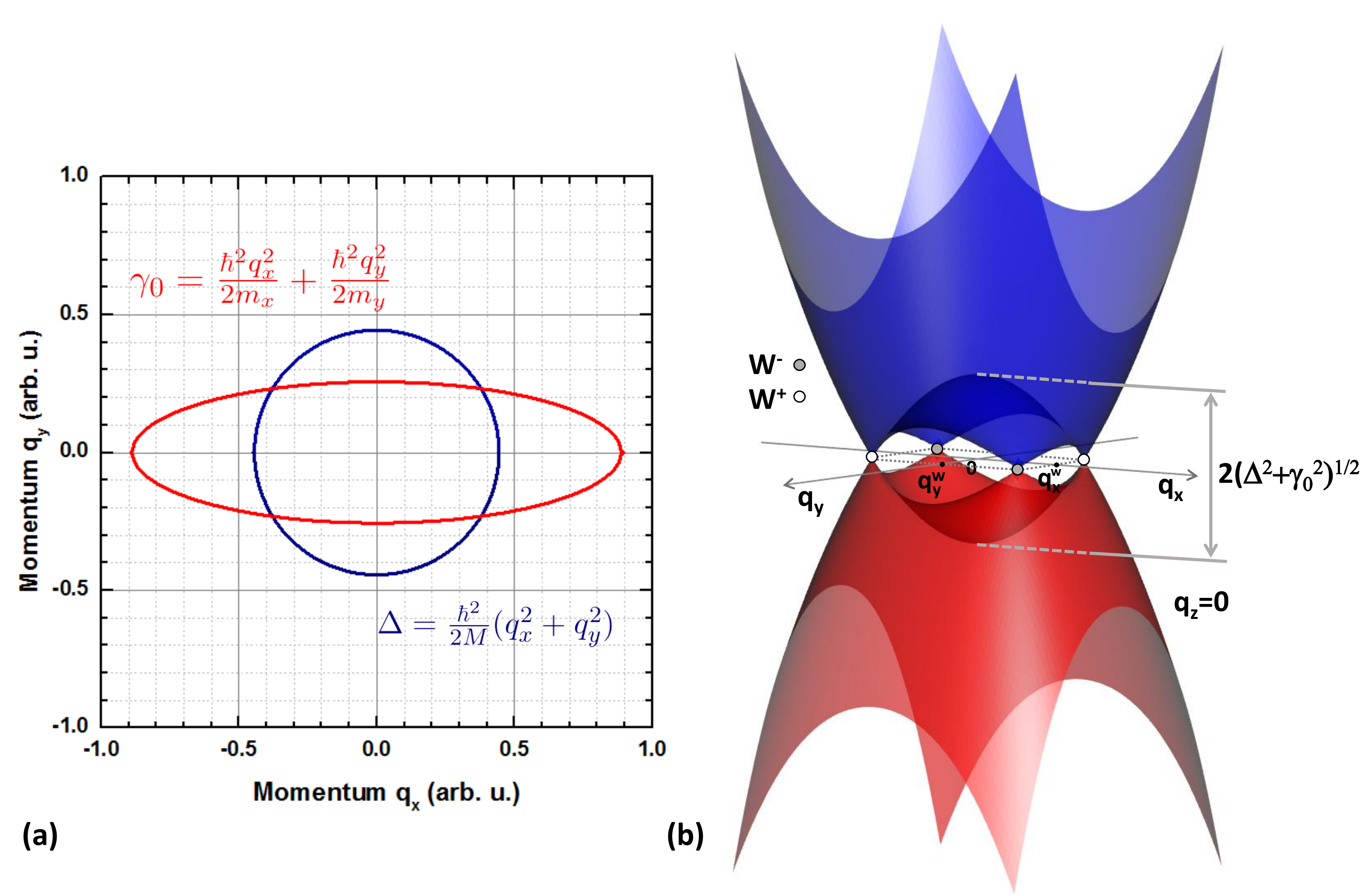}
      \caption{\label{Schematics} (a) The four zero-energy points of the Hamiltonian (\ref{hamiltonian_Mark}) appearing for a particular set of
      parameters at locations where both the diagonal and off-diagonal terms vanish. (b) The electronic band structure implied by the Hamiltonian (\ref{hamiltonian_Mark})
      with two pairs of Weyl cones located at $\mathbf{q}^w=(\pm q^w _x,\pm q^w_y,0)$ and full electron-hole symmetry. Two pairs of saddle points as well as one
      local extremum at $\mathbf{q}=0$ (due to the band inversion) appear in the conduction (blue) and valence (red) bands, respectively.}
\end{figure*}

In order to find a 3D Weyl semimetal that respects time-reversal symmetry, the coupling term must have the following non-isotropic form:
$\gamma(\mathbf{q})=\gamma_0-\hbar^2 q_x^2/(2m_x)-\hbar^2 q_y^2/(2m_y)+i \hbar v_z q_z$. For a particular choice of parameters,
four zero-energy points, $E(\mathbf{q}^w)=0$, appear at $\mathbf{q}^w=(\pm q_x^w, \pm q_y^w, 0)$.
For positive masses, $M,m_x,m_y>0$,
their positions may be determined as the crossing points of the circle, $\Delta = \hbar^2 (q_x^2+q_y^2)/(2M)$, with the ellipse, $\gamma_0=\hbar^2 q_x^2/(2m_x)+\hbar^2 q_y^2/(2m_y)$
defined by the diagonal and off-diagonal terms of (\ref{hamiltonian_Mark}), respectively (Fig.~\ref{Schematics}a).

At the zero-energy points, the conduction and valence bands meet and form four anisotropic, particle-hole-symmetric Weyl cones.
These cones merge at higher energies via two kinds of saddle points (Lifshitz transitions of neck-collapsing type~\cite{LifshitzZETF60,OrlitaPRL12}),
see Fig.~\ref{Schematics}b. Importantly, four is the
minimal number of nodes needed for a consistent model of a Weyl semimetal that respects time-reversal symmetry. Indeed, the Weyl
cones at opposite momenta \textbf{q}$^w$ and -\textbf{q}$^w$ have the same topological charge, and a second pair of Weyl cones is thus required to ensure
the sum of topological charges to be zero. The proposed model thus may be relevant for any time-reversal-invariant 3D Weyl semimetal. In the case of TaP,
the $\mathbf{q}=0$ point may be straightforwardly associated with the $\Sigma$ point
which is a time-reversal point at the edge of the first Brillouin, around which a closely-packed quartet of W$_1$ nodes is formed in the $q_z=0$ plane~\cite{WengPRX15,HuangNC15}.

The  simplicity of the model allows
us to find analytically the positions and topological charges of the Weyl nodes, as well as of the locations of the saddle points~\cite{SM}.
Simple expressions were also found for the energy of the saddle points, $E_{x(y)}^{\mathrm{sp}}=|M \Delta  -\gamma_0 m_{x(y)}|/\sqrt{M^2 + m_{x(y)}^2}$
at the $x$ ($y$) axis, and of the local conduction and valence band extrema, $E(\mathbf{q}=0)=\pm\sqrt{\Delta^2+\gamma_0^2}$ (Fig.~\ref{Schematics}b).
Furthermore, one may find the product of the effective velocities $v_+$ and $v_-$ of the Weyl cones,
$v_{\mathrm{eff}}^2=|v_+v_-|= q_x^wq_y^w\left(m_y^{-1}-m_y^{-1}\right)/M$,
which determines the cyclotron frequency of electrons in Weyl nodes, and thus also
the spectrum of low-energy Landau levels (LLs) discussed below~\cite{SM}.

\begin{figure*}[t]
      \includegraphics[width=.92\textwidth]{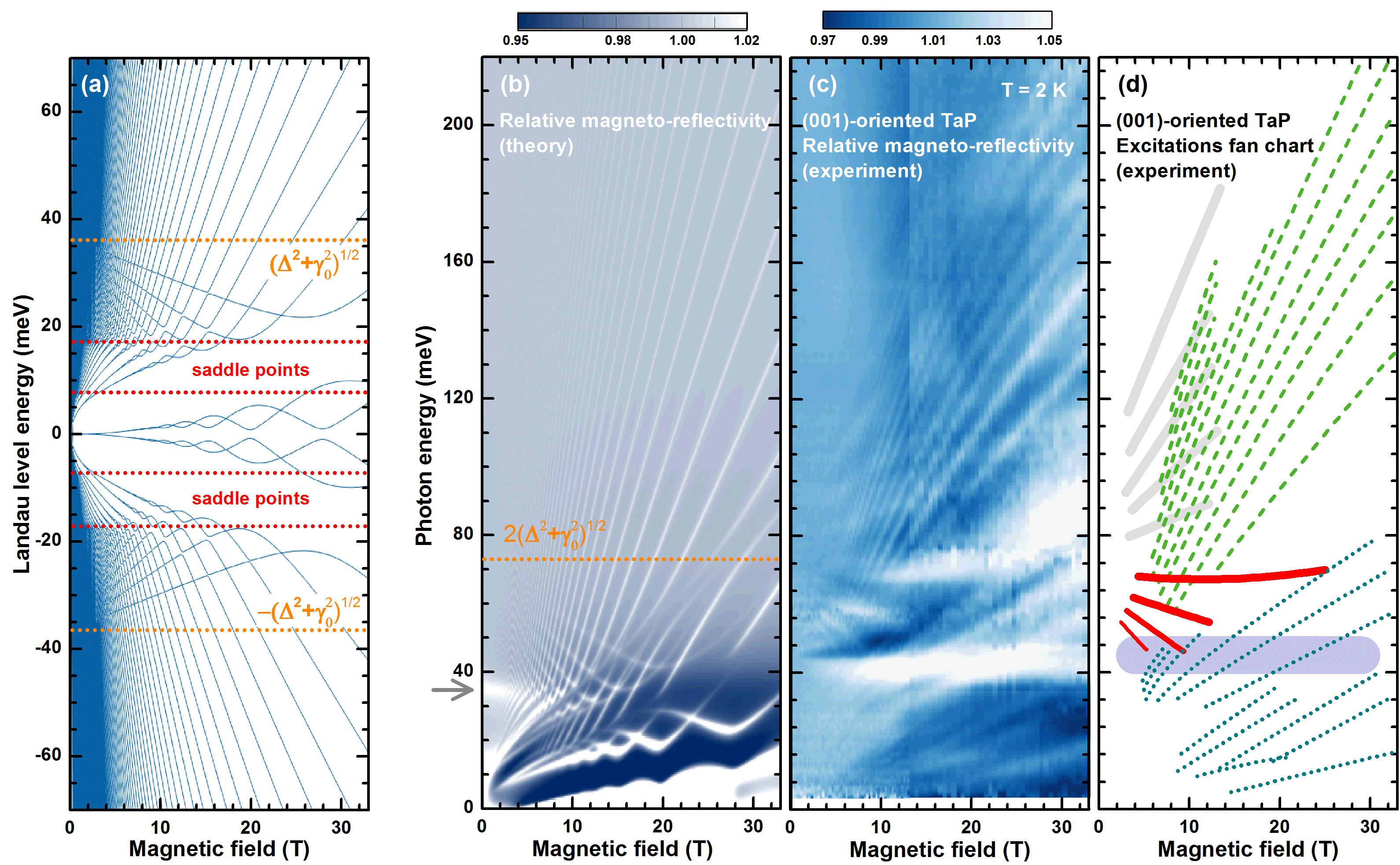}
      \caption{\label{Reflectivity}
(a) The LL spectrum as a function of $B$ calculated for $\Delta=35$~meV, $\gamma_0=10$~meV, $\Gamma=1$~meV and
$M=m_y=m_x/12=m_0/4$ where $m_0$ is the free electron mass. The number of LLs considered in our calculations has been limited to $N=700$.
The horizontal lines denote energies of the band extrema and saddle points in the (zero-field) band structure, cf. Fig.~\ref{Schematics}b. (b)
The false color-plot of relative magneto-reflectivity, $R_B/R_0$, calculated for the system considered in the panel (a).
The reference spectrum $R_0$ has been approximated by the reflectivity trace calculated at $B=1$~T and for $\Gamma=5$~meV. The background dielectric constant $\varepsilon_{\infty}=20$ has been considered. The horizontal arrow
indicates the feature due to excitations between higher-energy saddle points.
(c) The experimentally determined $R_B/R_0$ at $T=2$~K. (d) The schematic fan chart of
experimentally observed inter-LL excitations. For relatively narrow lines, maxima in $R_B/R_0$ represent good estimates of transition energies.
Excitations decreasing their energy with $B$ are represented by red solid lines.}
\end{figure*}

To test the model on our magneto-optical data, let us now introduce
a quantizing magnetic field along the $z$ axis in the Hamiltonian (\ref{hamiltonian_Mark}). We shall
consider only $q_z=0$ states which dominantly
contribute to the magneto-optical response (apart from specific effects related to linearly dispersing $n=0$ LLs~\cite{AshbyPRB13}).
Using the standard Peierls substitution and the representation
of ladder operators, $a = l_B(q_x + iq_y)/\sqrt{2}$, we obtain:
\begin{equation}
\hat{H}_B = \left( \begin{matrix}
\Delta - \hbar\omega_c\left(a^{\dagger}a+\frac{1}{2}\right) & \gamma(a,a^{\dagger}) \\
\gamma^*(a^{\dagger},a) & -\Delta + \hbar\omega_c\left(a^{\dagger}a+\frac{1}{2}\right)
\end{matrix} \right)
\label{hamiltonian_Mark_B}
\end{equation}
where $\omega_c$ denotes $eB/M$, $\gamma(a,a^{\dagger})$ stands for
$\gamma_0-(a+a^{\dagger})^2/(m_y l_B^2)-(a-a^{\dagger})^2/(m_x l_B^2)$, and $l_B=\sqrt{\hbar/(eB)}$ is the magnetic length.
Since the search for eigenvalues of  Hamiltonian (\ref{hamiltonian_Mark_B}) leads to diagonalization
of an infinite-size matrix, we limit ourselves to an approximate solution, taking account of a finite number of LLs ($N$)
and using the correspondingly truncated matrix.

An example of a calculated LL spectrum is presented in Fig.~\ref{Reflectivity}a.
At low energies, the LLs follow an almost perfect $\sqrt{B}$ dependence, characteristic for massless charge carriers~\cite{SadowskiPRL06},
and show a fourfold degeneracy. This degeneracy is gradually lifted with increasing $B$: a pair of doubly degenerate levels emerges and
subsequently, a quartet of non-degenerate LLs develops. This splitting occurs when the LLs approach the saddle points in the band structure
indicated by horizontal dashed lines in Fig.~\ref{Reflectivity}a.

In the intermediate energy range, the electron LLs exhibit rich (anti)crossing behavior. This happens above the saddle points, but still below
the local maximum of the conduction band, indicated by dotted lines in Fig.~\ref{Reflectivity}a.
This is due to specific conduction-band LLs, which peculiarly disperse towards lower energies with increasing $B$ due to inverted band ordering.
In the zero-field limit, they extrapolate to $\sqrt{\Delta^2+\gamma_0^2}$.
Due to electron-hole symmetry, an analogous set of LLs, which atypically for holes disperse towards higher energies with increasing $B$
and extrapolate to $-\sqrt{\Delta^2+\gamma_0^2}$,
also appears in the valence band. At high and low energies, all LLs follow nearly linear-in-$B$ dependence
which is dominantly governed by the diagonal terms in the Hamiltonian~(\ref{hamiltonian_Mark_B}).

Within linear-response theory, the magneto-optical conductivity for bulk materials and circularly polarized light reads:
\begin{equation}
\sigma^{\pm}(\omega,B)=2\frac{iG_0}{l_B^2 \omega}\sum_{n,m,q_z}\frac{(f_m-f_n)|\left<m|\hat{v}_{\pm}|n\right>|^2}{E_n-E_m-\hbar\omega+i\Gamma},
\label{Conductivity}
\end{equation}
where $G_0=e^2/(2\pi\hbar)$ is the quantum of conductance, $\Gamma$ the broadening
parameter, and $\hat{v}_{\pm}=\hat{v}_x\pm i \hat{v}_y$ are the velocity operators, with $\hbar\hat{v}_{x}=\partial\hat{H}/\partial q_x$ and
$\hbar \hat{v}_{y}=\partial\hat{H}/\partial q_y$. A closer analysis of the matrix elements, $\left<m|\hat{v}_{\pm}|n\right>$, shows that the
response is dominated by transitions following the selection rule $n=m\pm1$, typical of electron-dipole excitations in isotropic materials.
At the same time,
the pronounced anisotropy of the studied system also activates excitations beyond this selection rule, especially,
when the final or initial state is nearby the saddle point. The LL occupation follows the Fermi-Dirac distribution $f$ ($T=0$ in our calculations) and the LL indices
$m$ and $n$ run over all available initial and final states. Consistently with (\ref{hamiltonian_Mark_B}), we consider only
states around $q_z=0$. Therefore, we have replaced the sum over $q_z$ in (\ref{Conductivity}) by the integral of the characteristic $1/(E_m-E_n-\hbar\omega)^{1/2}$
profile in the joint density of states over the interval of ($E_m$-$E_n$; $E_m$-$E_n$+$\Gamma$).

Figure~\ref{Reflectivity}b displays a false-color plot of the relative magneto-reflectivity $R_B/R_0$ calculated theoretically from Eq. (\ref{Conductivity}),
for the same set of parameters as the LL spectrum in Fig.~\ref{Reflectivity}a. This map displays fairly complex behavior, nevertheless, one may still identify three
main sets of inter-LL excitations in Fig.~\ref{Reflectivity}b: (i) the transitions at low photon energies with $\sqrt{B}$, or more generally, sub-linear-in-$B$ dependence, which are related
to excitations within the Weyl cones, (ii) the transition between extrema of the inverted bands, which extrapolate to $2\sqrt{\Delta^2+\gamma_0^2}$ in the zero-field limit
and which disperse towards lower energies with increasing $B$, and (iii) a series of regularly-spaced and linear-in-$B$ excitations at high energies.

Let us now compare the theoretically calculated response with experiments (Fig.~\ref{Reflectivity}c) performed on the single TaP crystal, prepared by the method
described in Refs.~\onlinecite{ArnoldNC16,KimuraPRB17,BesserPRM19}. Our magneto-reflectivity data were taken in the Faraday configuration on the (001)-oriented facet.
A macroscopic area of the sample ($\sim4\:\mbox{mm}^2$), kept at $T=2$~K in the helium exchange gas and placed in a superconducting solenoid (below 13~T) or resistive coil (above 13~T), was exposed to the radiation of a globar or Hg lamp, which was analyzed by a Fourier-transform spectrometer and delivered via light-pipe optics. The reflected light was detected by a liquid-helium-cooled bolometer placed outside the magnet. The sample reflectivity $R_B$ at a given magnetic field $B$ was normalized by the sample's reflectivity $R_0$ measured at $B=0$.

Comparing the theoretically expected and experimentally measured response (Fig.~\ref{Reflectivity}b versus \ref{Reflectivity}c), reasonable semi-quantitative agreement is found.
One may identify all three groups of excitations in the experimental data
(dotted, dark-color solid and dashed lines in Fig.~\ref{Reflectivity}d, respectively). This agreement could be further improved by introducing electron-hole asymmetry and higher-order terms to the Hamiltonian (\ref{hamiltonian_Mark}). In fact, such expansion would be equivalent
to higher-order $k.p$ perturbation theory, which is often used, \emph{e.g.}, in semiconductors physics~\cite{YuFS96} and which would imply a number of additional material parameters.
Further improving of the quantitative agreement would thus be at the expense of the model's simplicity and of general applicability to time-reversal-invariant Weyls semimetals.

\begin{figure}[t]
      \includegraphics[width=.46\textwidth]{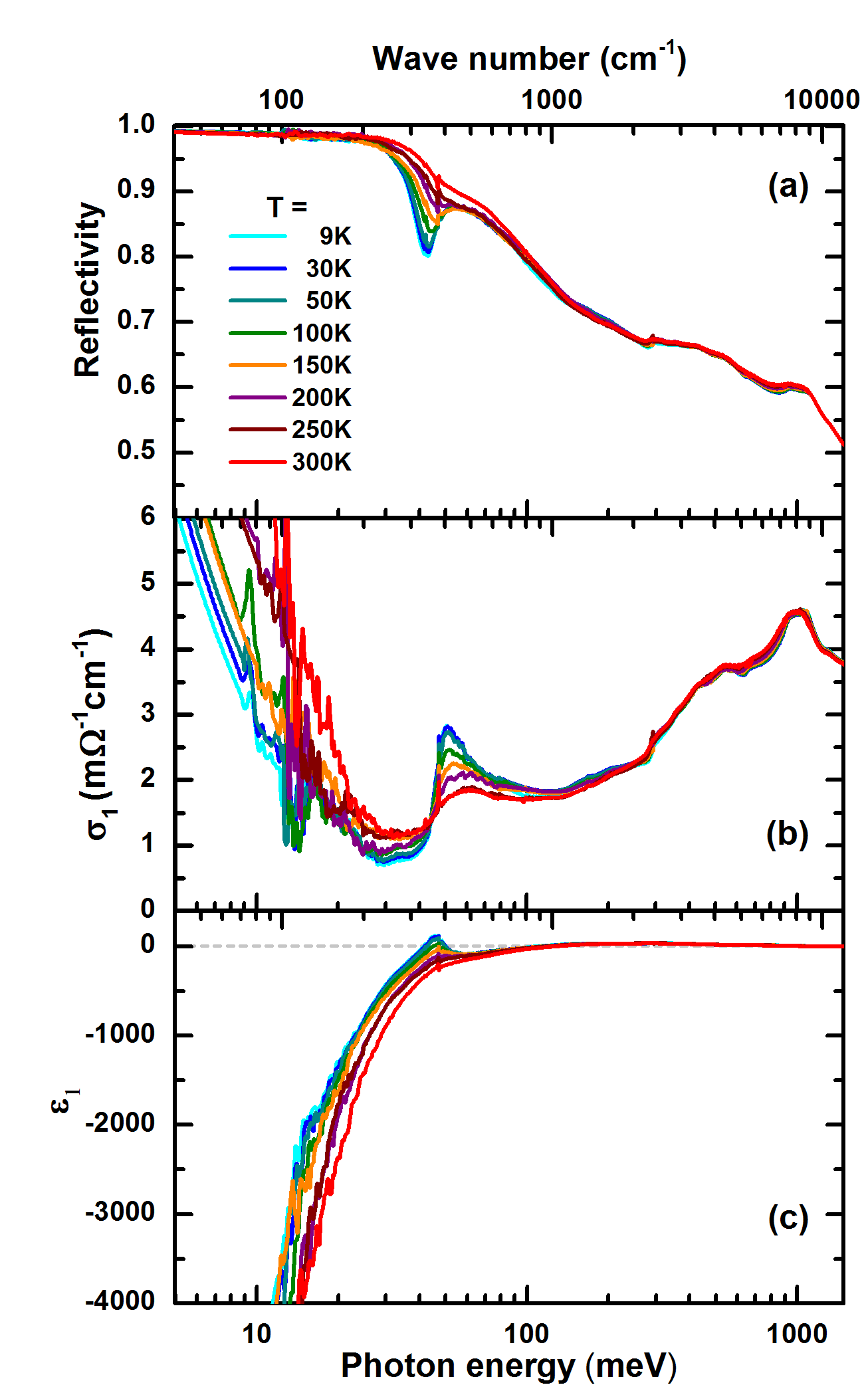}
      \caption{\label{Zero-field} (a) Infrared reflectivity of TaP measured at selected temperatures. The corresponding real part of conductivity
      and dielectric function are shown in parts (b) and (c), respectively. The measurements and analysis of this zero-field data
      closely followed the procedure described in Ref.~\onlinecite{NeubauerPRB18}.}
\end{figure}

Another similarity between the experiment and theory is linked to the maximum in the magneto-reflectivity data appearing around
$\hbar\omega \approx 40$~meV. It disperses only weakly with $B$ and thus becomes nearly horizontal in false color-plots in Figs.~\ref{Reflectivity}b,c.
In the theoretically expected response, this maximum (denoted by a horizontal arrow) appears due to excitations between
the saddle points. In the experimental data, however, the interpretation is less straightforward. The zero-field optical data (Fig.~\ref{Zero-field}) and
also the preceding optical study by Kimura~\emph{et al.}~\cite{KimuraPRB17}, indeed show in this spectral range a dissipative feature interpreted previously
as due to
excitations between the saddle points. Nevertheless, this feature nearly coincides in energy with the plasma edge (cf. Figs.~\ref{Zero-field}a, b and c),
which reflects the response due to all free electrons, including topologically trivial pockets. According to classical magneto-plasma theory~\cite{PalikRPP70},
the plasma edge in the reflectivity exhibits splitting with $B$ and may thus mask effects related to excitations
between saddle points.

The good agreement between theory and experiments calls for a more detailed discussion about the positions and energies of Weyl nodes in TaP. Ab initio
calculations~\cite{WengPRX15,LeePRB15,GrassanoSR18,GrassanoCM19}, also when confronted with ARPES, optical and transport
experiments~\cite{XuSA15,XuNatureComm16,ArnoldNC16,KimuraPRB17}, suggest that
there is a large energy separation between W$_1$ and W$_2$ nodes, $\Delta E_{\mathrm{W}_2-\mathrm{W}_1}\approx 60-80$~meV.
 Therefore, the Fermi level cannot be placed within both
types of cones at the same time and excitations in the intermediate range thus have to come from the same Weyl pocket~\cite{SM}.
In line with the model proposed above, we associate the observed response primarily with the W$_1$ pockets, with the cones in a closely packed quartet
in the vicinity of the Fermi level~\cite{SM}, rather than with fairly distant pairs of W$_2$ cones.
The latter pockets may only contribute, together with topologically trivial pockets, to low-energy
cyclotron resonances or at high energies, as interband excitations above the onset of $2\Delta E_{\mathrm{W}_2-\mathrm{W}_1}$. The latter were not identified
in the spectra.

To set appropriate parameters in the model, the results of ab initio studies have been used as a reference point~\cite{WengPRX15,GrassanoSR18,GrassanoCM19}. For the W$_1$ pocket,
these calculations imply the band inversion around $2\Delta_{\mathrm{DFT}}\approx 60$~meV at the $\Sigma$ point and the positions of nodes, $q_x^w\approx 0.4$~nm$^{-1}$ and
$q_y^w \approx 0.2$~nm$^{-1}$. This allow us to estimate the characteristic bending of the inverted conduction and valence bands described by the diagonal mass
$M=\hbar^2[(q_x^w)^2+(q_y^w)^2]/(2\Delta_{\mathrm{DFT}})\approx 0.25m_0$, as well as to set the off-diagonal masses: $m_y=m_x/12=m_0/4$. The
velocity parameter $v_z$, relevant in our approximation only for the strength of excitations, was $10^5$~m/s~\cite{GrassanoSR18}.
The comparison between theory and experiment, namely the zero-field extrapolation of transitions decreasing their energies with $B$, allowed us to fine tune
the band inversion, $2\sqrt{\Delta^2+\gamma_0^2}\approx 2\Delta \approx  73$~meV.
Notice that the coupling term $\gamma_0$ is thus the only tunable parameter of our model  (plausibly, $\gamma_0<\Delta$).
For simplicity, the Fermi level has been always kept at $E_F=0$.

The association of the observed response with the W$_1$ nodes located around the $\Sigma$ point 
allows us to explain
another set of excitations which are relatively weak, but still traceable in the experimental data and which are visualized by light-color solid lines in Fig.~\ref{Reflectivity}d.
The $\Sigma$ point at the edge of the Brillouin zone of TaP belongs to the time-reversal-invariant momenta where double (spin) degeneracy of the bands is ensured
by the Kramer's theorem. This implies the presence of two other electronic bands (not included in our simplified model and not plotted in Fig.~\ref{Schematics}b) which touch the
conduction and valence bands at $\mathbf{q}=0$. The additional set of transitions thus may be credibly assigned to interband excitations between such bands that may
also impact the LL spectrum at the $\Sigma$ point. This may result in a departure of our experiment from theory, especially for the spacing and slope of excitations decreasing
their energies with $B$.

In summary, we have proposed a simplified model for non-centrosymmetric Weyl semimetals to interpret high field magneto-reflectivity data.
In spite of the complexity of the TaP's band structure, our model explains reasonably well
our experimental observations. We also find that the unusual series of inter-LL excitations, which sweep towards lower energies with increasing $B$,
is directly related to the band inversion in TaP. In general, such behavior is rather rare, encountered only in a few experimental cases~\cite{SkolnickPRB94,HenniNL16},
mostly in the form of a single spectral line~\cite{ThilderkvistPRB94,MurdinNC13,RaymondPRL04,HeitmannPT93,LyonsNC19} rather than a full family of transitions as presented here.

We expect that similar series of excitations may be found in other time-reversal-invariant Weyl semimetals, and possibly also in a broader class of materials,
with inverted bands and weak coupling among them, such as nodal-loop semimetals with a vanishing band gap. Notice that this unusual set of inter-LL
excitations should not be typical of Weyl semimetals with broken time-reversal symmetry, where the conical bands appear in \textit{pairs} and the band inversion
does not necessarily lead to formation of a
well-defined local maximum and minimum in the band structure. Two cones, when merging via a single saddle point, do not give rise to any LLs dispersing negatively with
$B$~\cite{MontambauxEPJB09,deGailPRB11}. This is also seen in Fig.~\ref{Reflectivity}a, where the LL energies change monotonically with $B$
when passing through the saddle points, while their degeneracy is lifted.


\begin{acknowledgments}
The authors acknowledge helpful discussions with F. Bechstedt, S. Bordacs, O. Pulci, M. Potemski, A. O. Slobodeniuk. This work was supported by the ANR DIRAC3D project (ANR-17-CE30-0023). I.C. acknowledges funding from the Postdoc.Mobility fellowship of the Swiss National Science Foundation. The work was supported by the Deutsche Forschungsgesellschaft (DFG) via DR228/51-1. A.~A. acknowledges funding from the  Swiss National Science Foundation through project PP00P2\_170544.
\end{acknowledgments}


%

\newpage

\begin{figure}[htp]
\includegraphics[page=1,trim = 17mm 17mm 17mm 17mm, width=.99\textwidth,height=1.0\textheight]{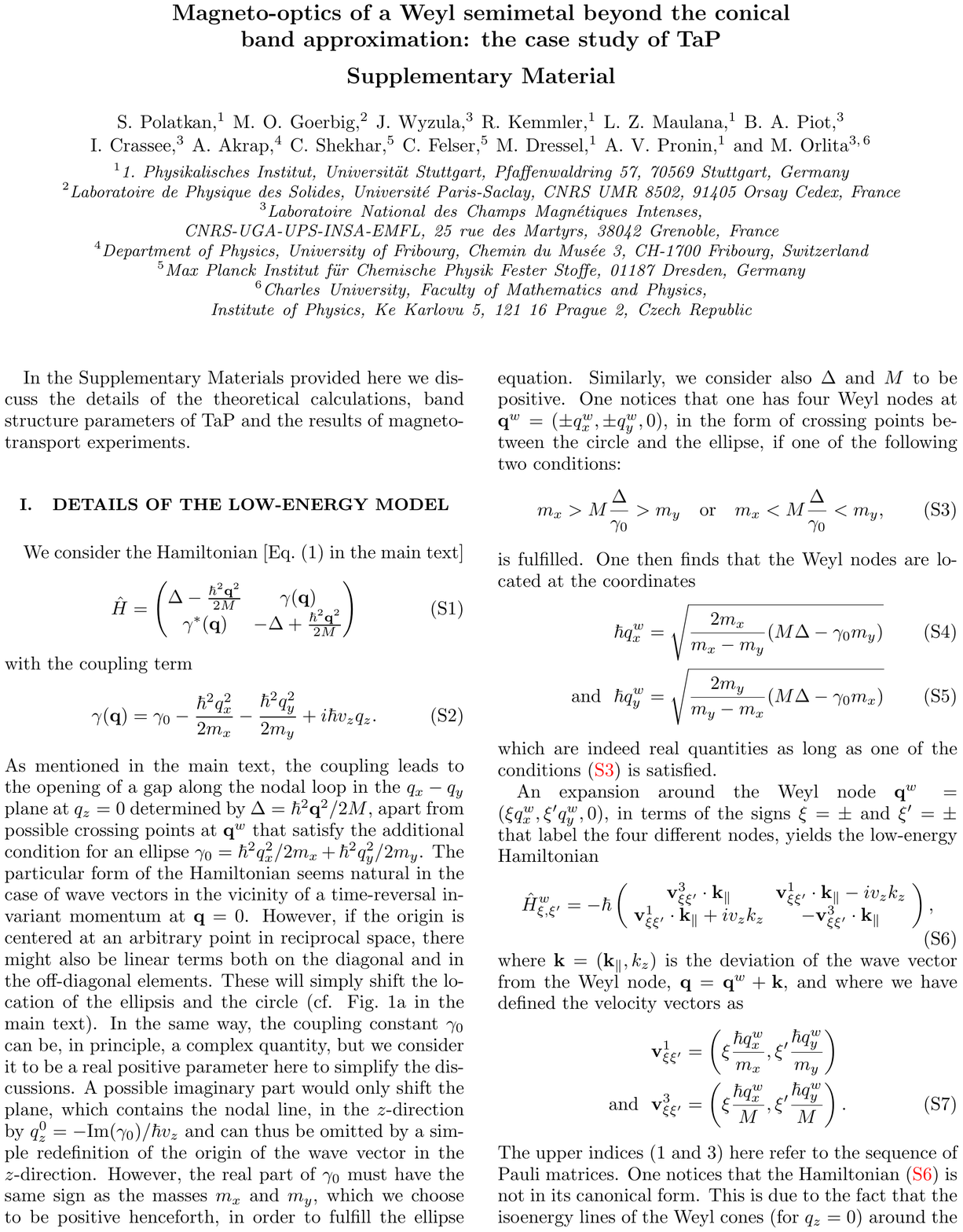}
\end{figure}

\newpage

\begin{figure}[htp]
  \includegraphics[page=2,trim = 17mm 17mm 17mm 17mm, width=.99\textwidth,height=1.0\textheight]{SM.pdf}
\end{figure}

\newpage

\begin{figure}[htp]
  \includegraphics[page=3,trim = 17mm 17mm 17mm 17mm, width=.99\textwidth,height=1.0\textheight]{SM.pdf}
\end{figure}

\newpage

\begin{figure}[htp]
  \includegraphics[page=4,trim = 17mm 17mm 17mm 17mm, width=.99\textwidth,height=1.0\textheight]{SM.pdf}
\end{figure}

\end{document}